\documentclass{ws-procs975x65}

\newcommand{\arcsecs}{\mbox{$^{\prime\prime}$}}

\begin{document}

\markboth{P. Jakobsson}{GRB-Selected Galaxies: A Swift/VLT Legacy Survey}


\title{FUNDAMENTAL PROPERTIES OF GRB-SELECTED GALAXIES: 
A SWIFT/VLT LEGACY SURVEY}

\author{P\'ALL JAKOBSSON}
\address{Centre for Astrophysics Research, University of Hertfordshire,
College Lane, \\ Hatfield, Herts, AL10 9AB, UK}

\author{JENS HJORTH and JOHAN P. U. FYNBO}
\address{Dark Cosmology Centre, Niels Bohr Institute, University of
Copenhagen, \\ Juliane Maries Vej 30, 2100 Copenhagen, Denmark}

\author{JAVIER GOROSABEL}
\address{Instituto de Astrof\'{\i}sica de Andaluc\'{\i}a (CSIC), 
Apartado de Correos 3004, 18080 Granada, Spain}

\author{ANDREAS O. JAUNSEN}
\address{Institute of Theoretical Astrophysics, PO Box 1029, 0315 Oslo, Norway}


\begin{abstract}
We present the motivation, aims and preliminary result from the
\emph{Swift}/VLT legacy survey on gamma-ray burst host galaxies.
This survey will produce a homogeneous and well-understood host 
sample covering more than 95\% of the lookback time to the Big Bang,
and allow us to characterize their fundamental properties.
\end{abstract}

\bodymatter

\section{Introduction}
With a very broad redshift distribution and a mean redshift of around
$z = 2.8$\cite{2.8}, gamma-ray bursts (GRBs) are becoming extremely
useful tracers of star-forming galaxies. Long-duration GRBs are known
to be associated with the deaths of short-lived massive stars\cite{JENS}
and thus have the essential advantage that their detection requires only 
a single stellar progenitor. Therefore, their detection is in principle 
independent of host galaxy luminosity.
\par
The \emph{Swift} satellite and a suite of ground-based observatories are 
detecting, localizing and studying a large homogeneous sample of GRBs. To 
take advantage of this unique sample, we have launched a dedicated
programme aimed at building up a sample of host galaxies, based on
\emph{Swift} detections and VLT follow-up. This is a Large Programme
to be executed over a period of two years. The resulting host sample will 
be largely unaffected by dust extinction and entirely independent of host
galaxy luminosity. A more thorough description of the survey and preliminary
results are presented in Hjorth et al. (in prep).
\par
The details of the sample selection are relatively straight-forward, i.e.
the GRBs have to be well-placed for optical follow-up observations: 
(i) Detected by \emph{Swift} after 1 March 2005 when it was fully operational
and automatically slewing. (ii) An X-ray position is available, obtained
by the \emph{Swift} XRT detector. (iii) The Galactic extinction is less
than $A_V < 0.5$\,mag. (iv) Declination favorable for VLT and not at a 
polar declination, i.e. $-70^\circ < \mathrm{dec} < 25^\circ$.
\section{Aims}
The concrete goals of the programme are to: (i) Identify the GRB hosts,
reaching a limit of around $R = 27.0$ and $K = 21.5$, which will allow
us to detect extremely red objects. For non-detections of hosts we will spend 
additional time to reach a limit of around $R = 28.0$. While hosts have been 
detected for nearly all pre-\emph{Swift} localized GRBs, almost none have 
been detected in the \emph{Swift} era. (ii) Measure redshifts for GRBs 
without absorption redshifts. (iii) Search for the Ly$\alpha$ emission line 
when possible, i.e. for bursts with a known redshift $z \gtrsim 2$. (iv) 
Study the effects of dust reddening within hosts. (v) Determine the host 
luminosity function. Finally, we will perform detailed studies of 
particularly interesting targets, e.g. short-duration GRB hosts and very 
bright hosts. Specifically, we will carry out emission line diagnostics,
e.g. metallicity estimates via the $R_{23}$ method.\cite{JAVIER}
\section{Results}
\begin{figure}[t]
\begin{center}
  \fbox{\epsfig{figure=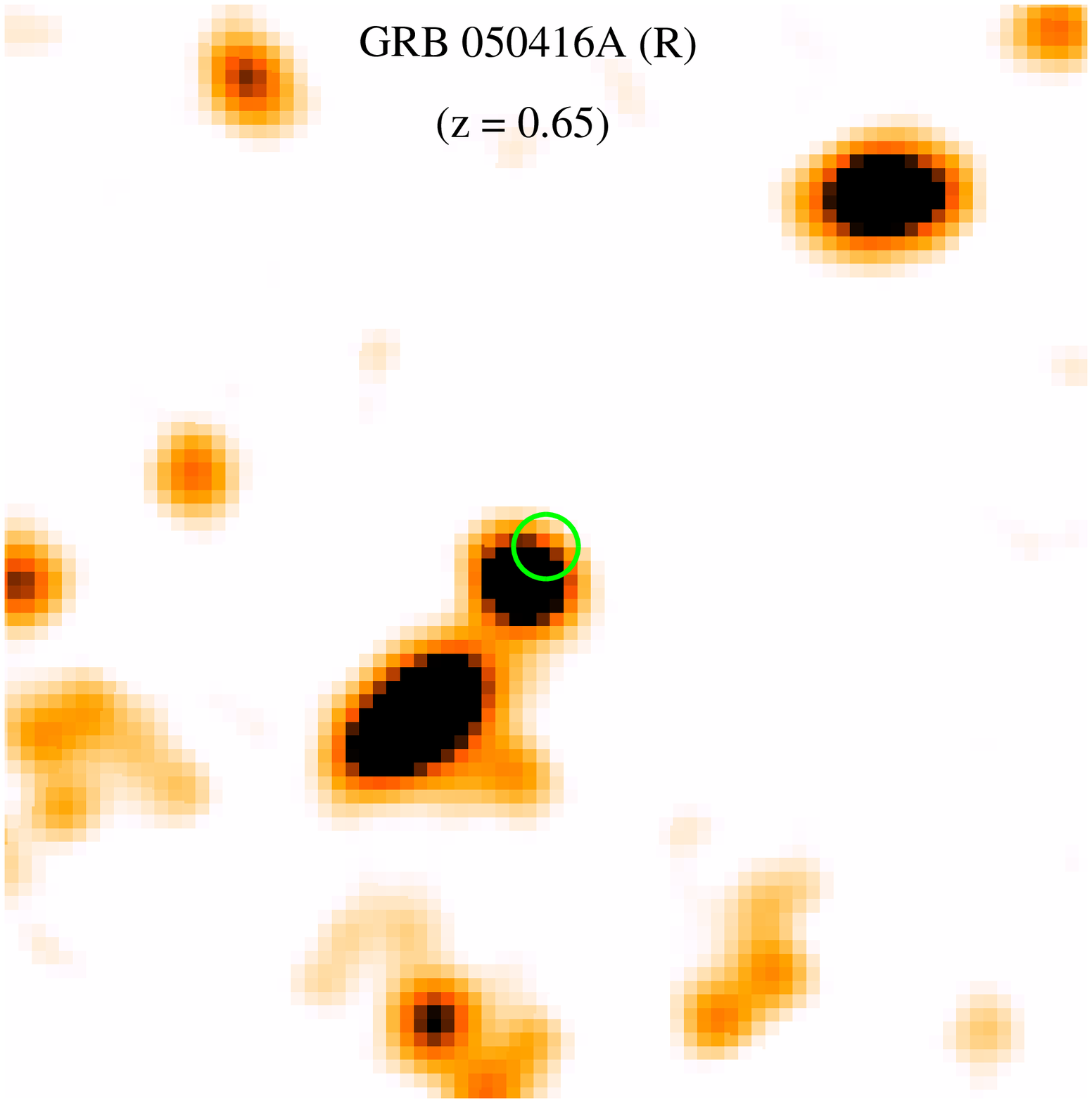,width=5.68cm}}
  \hspace*{-0.2cm}
  \fbox{\epsfig{figure=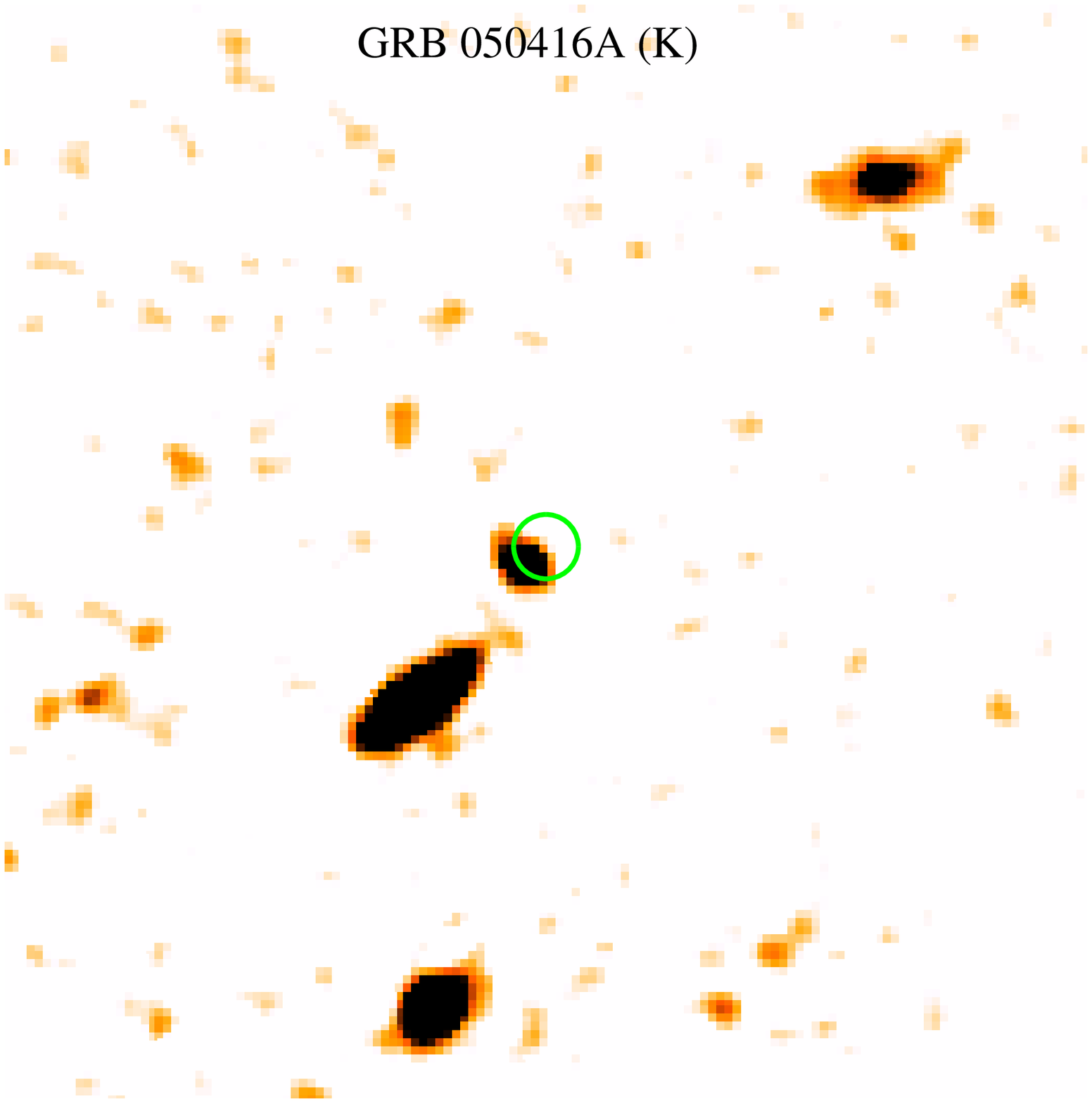,width=5.68cm}}
  \fbox{\epsfig{figure=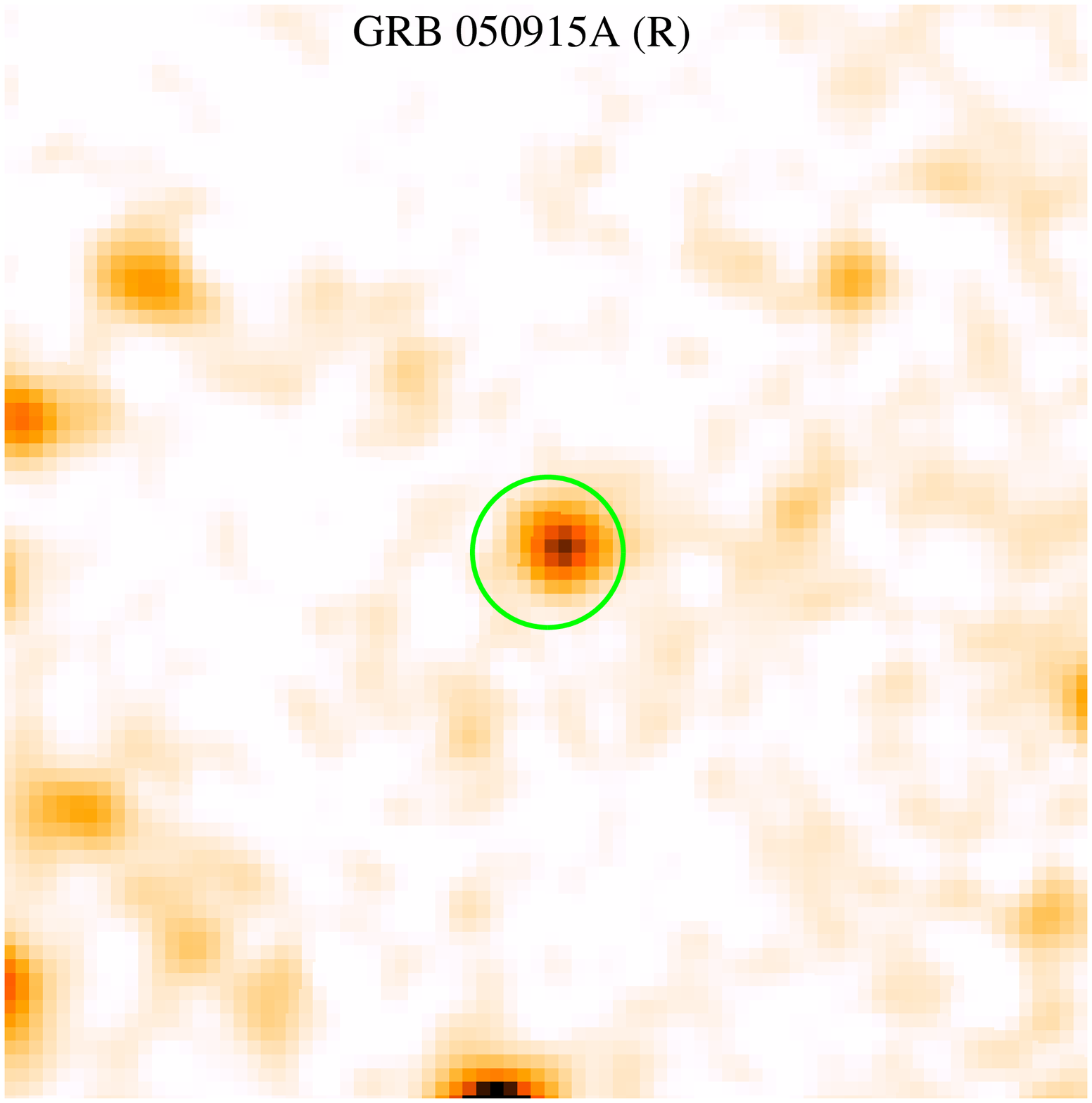,width=5.68cm}}
  \hspace*{-0.2cm}
  \fbox{\epsfig{figure=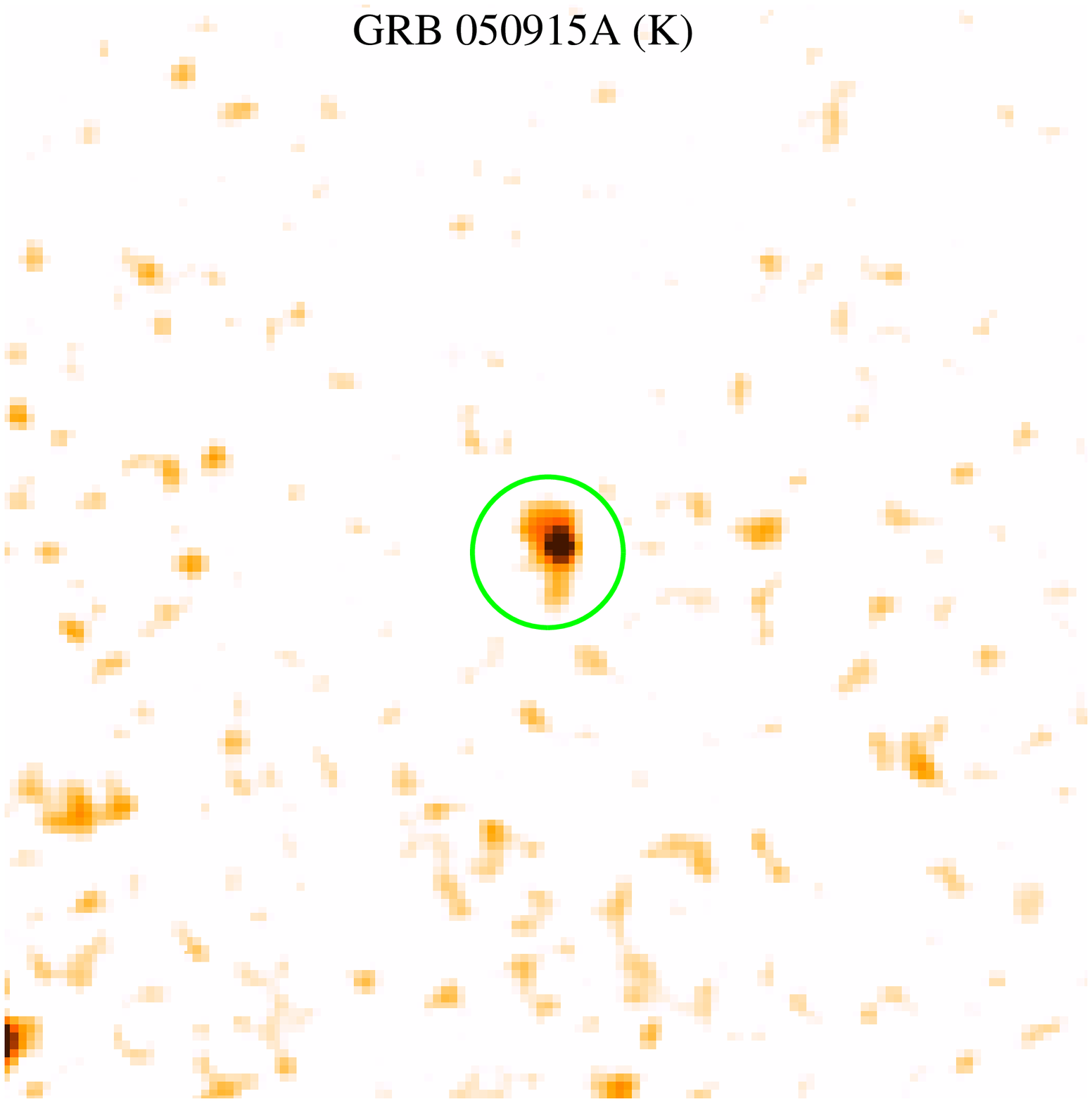,width=5.68cm}}
  \fbox{\epsfig{figure=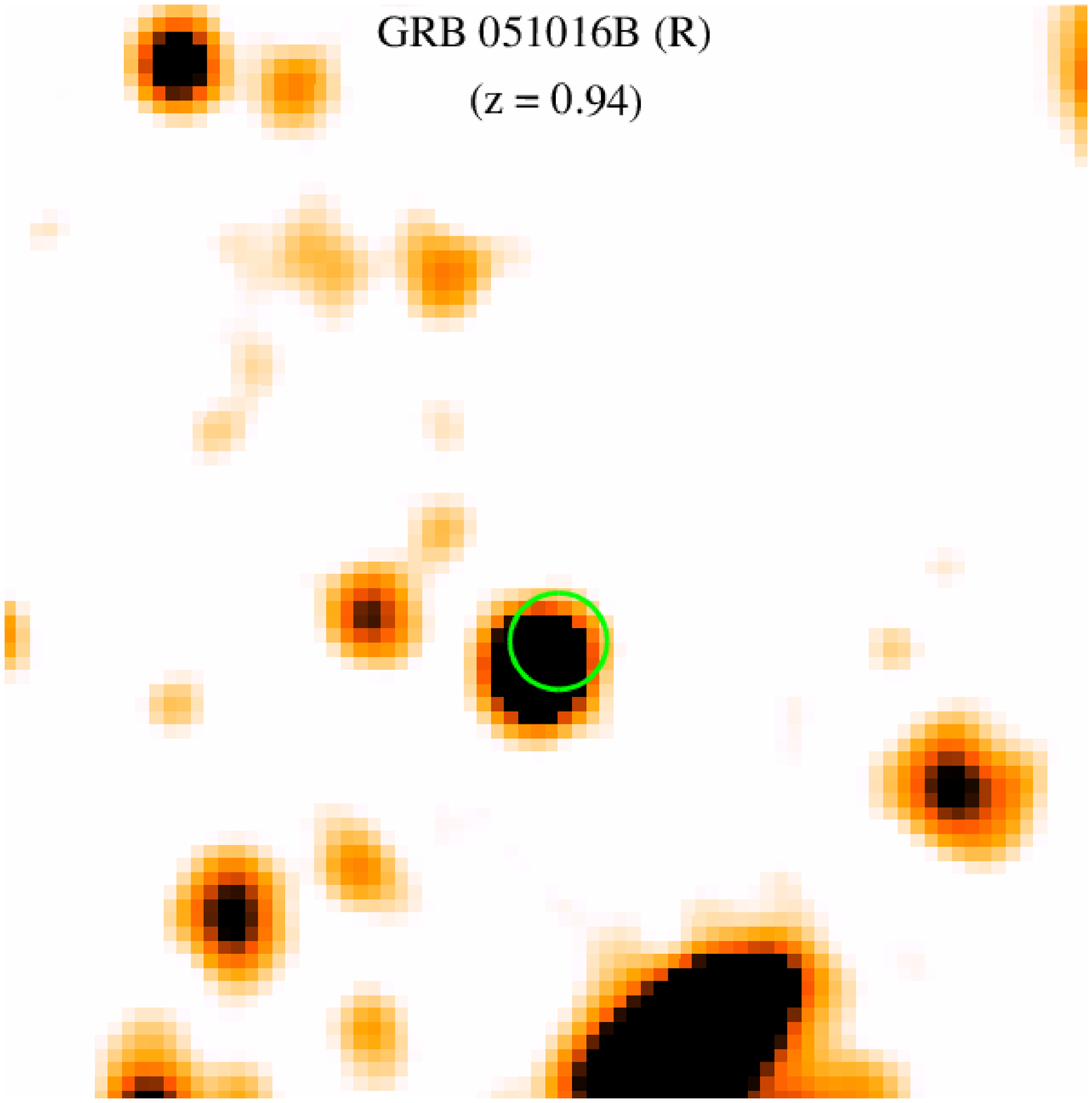,width=5.68cm}}
  \hspace*{-0.2cm}
  \fbox{\epsfig{figure=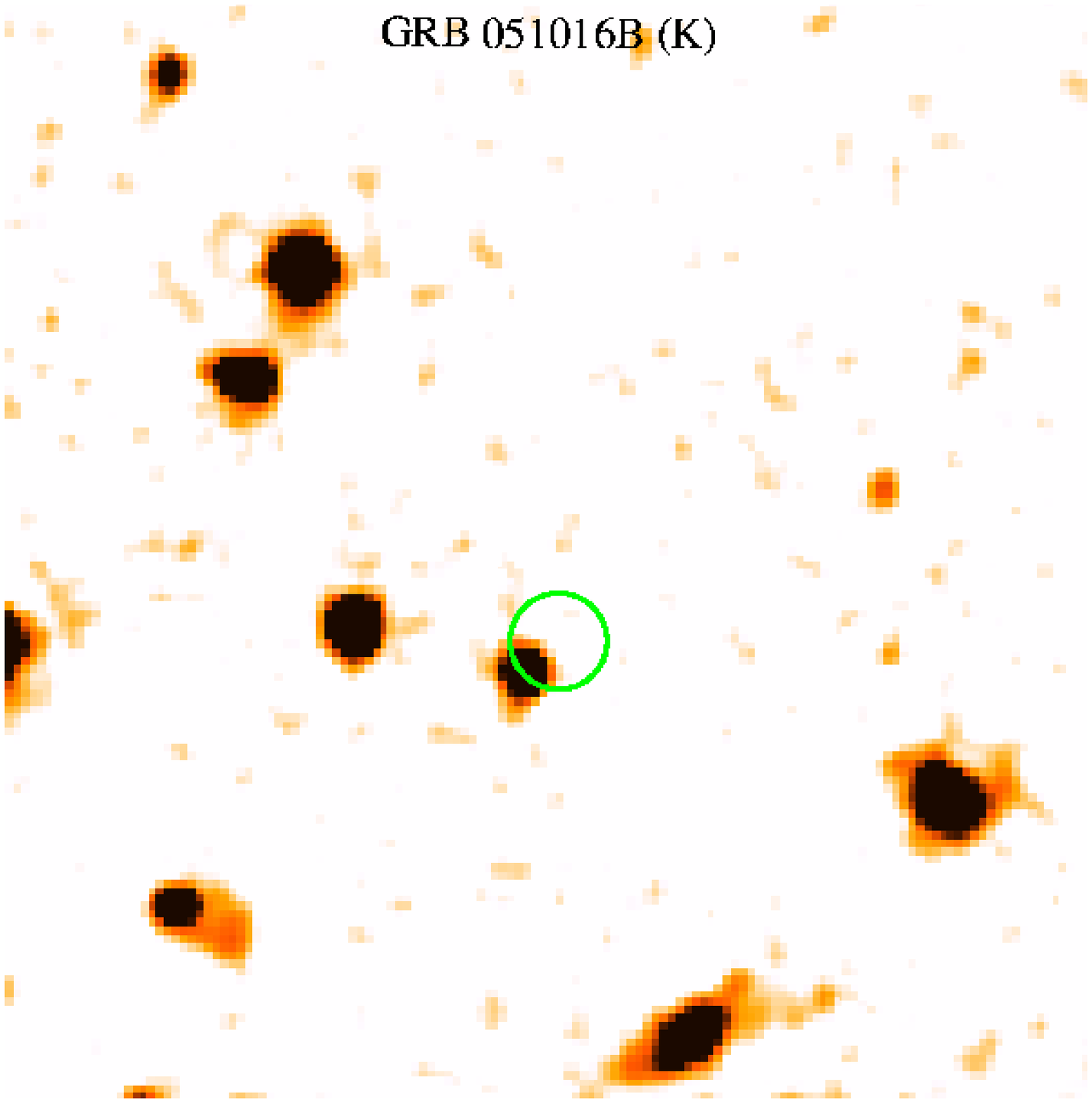,width=5.68cm}}
  \caption{A mosaic of three of the targets; left column displays the
  $R$-band while the $K$-band is in the right column. The host galaxy is
  detected in both bands for all targets, and is located inside the
  revised\cite{BUTLER} XRT error circle in each case (solid circle). Each
  host galaxy also coincides with the corresponding optical afterglow.
  The GRB 050915A host and all the $K$-band host detections have not been 
  reported before. North is up and east left in each panel which is 
  20\arcsecs\ on a side.}
  \label{hosts.fig}
\end{center}
\end{figure}

The final host sample is expected to consist of approximately 70 galaxies
of which a major fraction will have redshifts. The programme so far has
consisted mostly of target build-up, observational preparation, data taking
and preliminary analysis. To date, only six months after the start of the
programme, we have completed roughly half of item (i) above; $R$- and $K$-band 
imaging of three of the hosts is displayed in Fig.~\ref{hosts.fig} as an 
example. The current average and median $R$-band magnitude of the sample is 
fainter than 25.5.
\par
With this programme, we hope to detect a number of faint galaxies (such
as the GRB 030323 host\cite{PAUL}) that possibly dominate\cite{PALLI} the
total star-formation density at $z \gtrsim 2$, but are impossible to find
and study by other methods than GRB selection. But most importantly, we
will produce a coherent sample of GRB host galaxies for future follow-up
with the \emph{HST}, \emph{Spitzer}, VLT, and later with ALMA and
\emph{JWST}.

\vfill

\end{document}